\documentclass[english,reprint, aps, prl,superscriptaddress, showpacs,
showkeys, longbibliography, amsmath, amssymb, floatfix]{revtex4-1} 
\pdfoutput=1

\usepackage{fullpage}
\usepackage{cmap}
\usepackage[T1]{fontenc}
\usepackage[utf8]{inputenc}
\usepackage[french,main=english]{babel}
\usepackage{amsthm}
\usepackage{mathrsfs}
\usepackage{bbold}
\usepackage{wesa}
\usepackage{graphicx}
\usepackage{verbatim}
\usepackage[backref=false]{hyperref}
\usepackage{booktabs}
\usepackage{multirow}
\usepackage[braket,qm]{qcircuit}
\usepackage{color}
\usepackage[usenames,dvipsnames]{xcolor}
\usepackage{framed}
\usepackage{comment}
\usepackage{xparse}

\theoremstyle{plain}
\newtheorem{thm}{Theorem}
\newtheorem{lemma}{Lemma}

\theoremstyle{definition}
\newtheorem{definition}{Definition}

\DeclareUnicodeCharacter{0229}{\c{e}}

\newcommand{\Hilb}{\mathcal{H}}
\newcommand{\events}{\ensuremath{\mathcal{E}}}

\newcommand{\interval}[1]{{\normalfont\textsf{\textbf{#1}}}}
\newcommand{\imposs}{\interval{F}}
\newcommand{\necess}{\interval{T}}
\newcommand{\unknown}{\interval{U}}

\newcommand{\proj}[1]{\op{#1}{#1}}
\newcommand{\ps}{\texttt{+}}

\newcommand{\set}[2]{\ensuremath{\left\{ {#1}\mathrel{}\middle|\mathrel{}{#2}\right\} }}
\newcommand{\Tr}{\ensuremath{\mathop{\mathrm{Tr}}\nolimits}}
\allowdisplaybreaks
\newcommand{\coreBorn}{\ensuremath{\overline{\Hilb}}}

\newcommand{\rmi}{i}
\newcommand{\ultramodular}{\mathcal{M}}
\newcommand{\ultramodularL}[1][]{\ensuremath{\ultramodular^{L{#1}}}}
\newcommand{\ultramodularR}[1][]{\ensuremath{\ultramodular^{R{#1}}}}
\newcommand{\muB}{\ensuremath{\mu^{B}}}
\newcommand{\eventsC}{\ensuremath{\events_{C}}}

\edef\docIndent{\the\parindent}
\edef\docParskip{\the\parskip}
\newenvironment{docParinfo}{%
 \edef\currentIndent{\the\parindent}
 \edef\currentParskip{\the\parskip}
 \setlength{\parindent}{\docIndent}%
 \setlength{\parskip}{\docParskip}
}{
 \setlength{\parindent}{\currentIndent}%
 \setlength{\parskip}{\currentParskip}
}
\newcommand{\says}[3]{\begin{framed}\begin{minipage}{0.9\linewidth}\color{#1}{#2 says: #3}\end{minipage}\end{framed}}

\newcommand{\yutsung}[1]{\says{purple}{Yu-Tsung}{#1}}

\newcommand{\andy}[1]{\says{blue}{Andy}{#1}}
\NewDocumentEnvironment{suggest}{m m O{}}{#1{TEXT \##2 would go here. #3}}{#1{TEXT \##2 end here.}}
\NewDocumentEnvironment{compare}{m}{\begin{widetext}#1{The following left column should be replaced by the right column.}\end{widetext}}{\begin{widetext}#1{End of replacement.}\end{widetext}}
\newenvironment{compareText}{\noindent\begin{minipage}[b]{1\columnwidth}\begin{docParinfo}}{\end{docParinfo}\end{minipage}}

\begin{document}

\title{Quantum Interval-Valued Probability: Contextuality and the Born Rule}

\author{Yu-Tsung Tai}
\affiliation{Department of Mathematics, Indiana University, Bloomington, Indiana 47405,
USA}
\affiliation{Department of Computer Science, Indiana University, Bloomington,
Indiana 47408, USA}

\author{Andrew J. Hanson}
\affiliation{Department of Informatics, Indiana University, Bloomington,
Indiana 47408, USA}

\author{Gerardo Ortiz}
\affiliation{Department of Physics, Indiana University, Bloomington, Indiana 47405,
USA}

\author{Amr Sabry}
\affiliation{Department of Computer Science, Indiana University, Bloomington,
Indiana 47408, USA}

\date{\today}

\begin{abstract}
  We present a mathematical framework based on quantum interval-valued
  probability measures to study the effect of experimental
  imperfections and finite precision measurements on defining aspects
  of quantum mechanics such as contextuality and the Born rule. While
  foundational results such as the Kochen-Specker and Gleason theorems
  are valid in the context of infinite precision, they fail to hold in
  general in a world with limited resources. Here we employ an
  interval-valued framework to establish bounds on the validity of
  those theorems in realistic experimental environments. In this way,
  not only can we quantify the idea of finite-precision measurement
  within our theory, but we can also suggest a possible resolution of
  the Meyer-Mermin debate on the impact of finite-precision
  measurement on the Kochen-Specker theorem.
\end{abstract}



\maketitle

\section{Introduction}

In this investigation, we explore the implications of extending
conventional quantum-mechanical probability measures to include the
effect of imperfect measurements.  There are long-standing debates in
the foundations of quantum mechanics regarding the tension between
finite precision measurement and
contextuality~\cite{BarrettKent2004,Appleby_2005}. The outcome of a
theory that is contextual depends upon whether compatible sets of
observables are measured together or separately; a non-contextual
theory gives the same results in either case.  A classic example of
the varying opinions on the impact of imprecision is the claim by
Meyer~\cite{PhysRevLett.83.3751} that finite-precision measurements
invalidate the spirit of the Kochen-Specker theorem. This claim was
countered in the same year by Havlicek et al.~\cite{HKSS1999apsrev4}
and Mermin~\cite{Mermin1999} and the debate continues to be an active
topic of
research~\cite{Kent1999,SimonBruknerZeilinger2001,Cabello2002,Larsson2002,Appleby2002,BarrettKent2004,Appleby_2005,Spekkens2005,GuehneKleinmannCabelloEtAl2010,MazurekPuseyKunjwalEtAl2016}.

The Kochen-Specker
theorem~\cite{BELL_1966,kochenspecker1967,Redhead1987-REDINA,Mermin1990Simple,peres1995quantum,Jaeger2007,Held2016}
is in essence a mathematical statement about contextuality, asserting
that in a Hilbert space of dimension $d \ge 3$, it is impossible to
associate determinate probabilities, $\mu(P_i)= 0$ or $1$, with every
projection operator~$P_i$, in such a way that, if a set of commuting
$P_i$ satisfies $\sum_i P_i = \mathbb{1}$, then $\sum_i \mu(P_i) = 1$.
Meyer showed that one {\it can\/} assign determinate probabilities $0$
or~$1$ to the measurement outcomes when the Hilbert space is defined
over the field of rational numbers, therefore nullifying the
theorem~\cite{PhysRevLett.83.3751}.  However, Mermin then argued that
measurement outcomes must depend smoothly on slight changes in the
experimental configuration, leading him to assert that the impact of
Meyer's negative result is ``unsupportable'' on physical
grounds~\cite{Mermin1999}.  The question of how to address this controversy, and the effect of 
finite precision on measurements in general, is therefore 
a fundamental problem of physics.  How does one
develop mathematical theories of quantum mechanics that are
intrinsically, rather than only implicitly, consistent with the
resources available for the realistic accuracy of an actual
measurement?   

In this paper, we extend our previous work on the
foundations of computability in quantum physics by exploring the
application of interval-valued probability measures (IVPMs) to
achieving a coherent formulation of finite-resource quantum mechanics.
From this starting point, we develop a mathematical framework that includes the
uncertainties of finite precision and imperfections in the quantum
measurement process.  In particular, we are able to suggest a way to quantify
the concept of imperfect quantum measurement and its impact on the interpretation of
the Kochen-Specker and Gleason theorems and their implications for the
foundations of quantum mechanics.  We have reason to believe that
essential objectives of our program to achieve a computable theory of
quantum mechanics and finite-resource measurement, attempted
previously with computable number
systems~\cite{usat,geometry2013apsrev4,DQT2014}, may be achievable by
extending classical IVPMs~\cite{JamisonLodwick2004} to the quantum
domain.  We thus begin by axiomatizing a quantum interval-valued
probability measure (QIVPM) framework.

To put our investigation in perspective, we note that Meyer attributes
finite-precision errors \emph{exclusively} to the description of the
states defined in a general Hilbert space. Mermin's response argues
against Meyer's interpretation but, indeed,  accepts his framework. Others,
however, have argued that the effect of finite-precision measurements
on the Kochen-Specker theorem requires a different approach altogether
(see for example Ax and Kochen's communication cited by
Cabello~\cite{Cabello2002}). Here, we propose such a new approach. 

Our approach, based on QIVPMs, introduces the concept of $\delta$-determinism,
where~$\delta$ quantifies the effect of finite-precision uncertainties
on measurement outcomes.  Our parameter $\delta$ is not necessarily
related to the description of states; we are agnostic about
attributing elements of reality to the state function. Instead, we
attribute lack of certainty to the uncertain results obtained through
the use of a measuring instrument.  Thus, the parameter~$\delta$ in
our framework reflects insufficient knowledge of the experimenter,
which could be due to a variety of reasons related to imperfections of
devices. For example, a typical question that an experimenter may not
be able to answer accurately would be ``did an electron land in the
left half or the right half of the screen?''  There are cases in which
the electron would land too close to the middle of the screen for the
experimenter to be able to determine with certainty that it was left
or right of the center.  We simply record this imprecision by
generalizing the probabilities assigned to events to intervals reflecting the
uncertainties.


We then recast the Kochen-Specker theorem using
$\delta$-determinism to quantify the effects of finite precision and
imperfections on contextuality. When $\delta=0$, the generalized
theorem reduces to the conventional one, but as $\delta$ varies from 0
to 1 we note a transition from contextuality to non-contextuality. For
a \emph{non-vanishing} range of~$\delta$, quantum-mechanical
contextuality continues to hold, maintaining the Kochen-Specker
result, but at a \emph{certain fixed value}, $\delta = \frac{1}{3}$,
there is a sharp transition to \emph{non-contextuality}, parallel in
spirit to a phase transition.  These results provide a new insight
into the Meyer-Mermin debate by presenting a theory in which there is
a \emph{parameter} interpolating between what appeared previously to
be irreconcilable aspects of contextuality.

We also investigate the second key aspect of imprecise quantum
measurement, which is its statistical nature as reflected in the
application of the Born rule \cite{Born1983bibTeX,Mermin2007,Jaeger2007} determining the probability of a given
measurement outcome.
Here, Gleason's theorem \cite{gleason1957,Redhead1987-REDINA,peres1995quantum} establishes that there is no alternative to
the Born rule by demonstrating, using reasonable continuity arguments
in Hilbert spaces of dimension $d\ge3$, the existence of a unique
state $\rho$ consistent with the statistical predictions computed from
the Born rule.
Our question then concerns what happens when infinite precision cannot
be achieved.  We are able to show that, while a QIVPM incorporating
the effects of finite precision might not be consistent with Gleason's
unique state~$\rho$ on all projectors defined on a Hilbert
space~$\Hilb$ of dimension $d\ge3$, there is a mathematically precise
sense in which one recovers the original Gleason theorem
asymptotically.  Specifically, it is possible to construct a class of
QIVPMs representing bounded resources that is parameterized by the
size of the intervals. We then demonstrate that all QIVPMs in this
class are consistent with a non-empty ``ball'' of quantum states whose
radius is defined by the maximal length of the intervals
characterizing the uncertainties. As the size of the intervals goes to
zero, this ball of quantum states converges to a point representing
the unique state consistent with the Born rule and Gleason's theorem.

The paper is organized as follows.  In Sec.~\ref{sec:fuzzy} we begin
by introducing the foundations of quantum probability space and fuzzy
measurement.  We then move on to introduce the concept of quantum
interval-valued probability measures in
Sec.~\ref{sec:Interval-Uncertainty}.  These QIVPMs will provide the
mathematical framework we need to quantify the impact of finite
precision measurements on quantum mechanics.  For instance, in
Sec.~\ref{sec:Kochen-Specker}, we quantify the domain of validity of a
contextual measurement, thus addressing the conditions under which the
Kochen-Specker theorem applies.  This provides a way not only of
resolving the Meyer-Mermin debate, but also of revealing a precise
transition to non-contextuality; these are the results of our
Thm.~\ref{cor:Kochen-Specker-IVPM}.  In Sec.~\ref{sec:Gleason}, we
study Gleason's theorem in the context of imprecise measurements,
concluding that QIVPMs provide a framework with quantitative bounds in
which Gleason's theorem, while formally invalid in a universe with
bounded resources, holds asymptotically in a mathematically precise
way.  Finally, we present our conclusions in
Sec.~\ref{sec:Conclusion}.

\section{Fuzzy Measurements}
\label{sec:fuzzy}

A \emph{probability space} is a mathematical abstraction specifying
the necessary conditions for reasoning coherently about collections of
uncertain
events~\cite{Kolmogorov1950bibTeX,544199,Griffiths2003,Grabisch2016}. In the
quantum case, the events of interest are specified by \emph{projection
  operators} $P$ satisfying the condition $P^2=P$. These include the
empty projector~$\mathbb{0}$, the identity projector~$\mathbb{1}$,
projectors of the form $\proj{\phi}$ where $\ket{\phi}$ is a pure
quantum state (an element of a Hilbert space ${\cal H}$), sums of
\emph{orthogonal} projectors~$P_0$ and~$P_1$ with $P_0P_1=\mathbb{0}$,
and products of \emph{commuting} projectors $P_0$ and~$P_1$ with
$P_0P_1=P_1P_0$. In a quantum probability
space~\cite{10.2307/2308516,gleason1957,Redhead1987-REDINA,Maassen2010,Abramsky2012},
each event~$P_{i}$ is mapped to a probability~$\mu(P_{i})$ using a
probability measure~$\mu:\events\rightarrow[0,1]$, where $\events$ is
the set of all events, (i.e., projectors on a given Hilbert space), 
subject to the 
following constraints: $\mu(\mathbb{0})=0$, $\mu(\mathbb{1})=1$,
$\mu\left(\mathbb{1}-P\right)=1-\mu\left(P\right)$, and for each pair
of \emph{orthogonal} projectors $P_{0}$ and $P_{1}$:
\begin{equation}
{\mu}\left(P_{0}+P_{1}\right)={\mu}\left(P_{0}\right)+{\mu}\left(P_{1}\right)\,.\label{eq:QuantumProbability-Addition}
\end{equation}
Given a Hilbert space ${\cal H}$ of dimension $d$ and a probability assignment
for every projector $P$, we can define the expectation value of an
observable~$\mathbf{O}$ having spectral decomposition
$\mathbf{O}=\sum_{i=1}^{d}\lambda_{i}P_{i}$, with eigenvalues $\lambda_i \in
\mathbb{R}$, as~\cite{544199,Jaeger2007}:
\begin{equation}
\expval{\mathbf{O}}_{\mu}=\sum_{i=1}^{d}\lambda_{i}\mu(P_{i})\,.\label{eq:quantum-expectation}
\end{equation}

A conventional quantum probability measure can easily be constructed
using the Born rule if one knows the current pure normalized quantum
state $\ket{\phi} \in {\cal H}$; then the Born rule induces a
probability measure $\muB_{\phi}$ defined as
$\muB_{\phi}(P)=\melem{\phi}{P}{\phi}$. For mixed states
$\rho = \sum_{j=1}^{N}q_{j}\proj{\phi_{j}}$, where
$\ket{\phi_j} \in {\cal H}$, $q_j > 0$, and $\sum_{j=1}^{N}q_{j}=1$,
the generalized Born rule induces a probability measure $\muB_{\rho}$
defined as
$\muB_{\rho}\left(P\right) = \Tr\left(\rho P\right) = \sum_{j=1}^{N}
q_{j}\muB_{\phi_{j}}\left(P\right)$~\cite{Born1983bibTeX,Mermin2007,Jaeger2007}.

The quantum probability postulates assume a mathematical idealization
in which quantum states and measurements are both infinitely
precise, i.e., {\it sharp}. 
In an actual experimental setup with an ensemble of quantum
states that would ideally be identical, but are not actually
identically prepared, with imperfections and inaccuracies in measuring
devices, an experimenter might only be able to determine that the
probability of an event $P$ is concentrated in the range $[0.49,0.51]$
instead of being precisely $0.5$. The spread in this range depends on
the amount of resources (time, energy, money, etc.)  that are devoted
to the experiment. In the classical setting, this ``fuzziness'' can be
formalized by moving to \emph{interval-valued probability measures}
(IVPMs), which we explore in the next section, along with our proposed
extension to the quantum domain.

\section{Intervals of Uncertainty}
\label{sec:Interval-Uncertainty}

We will start by reviewing classical IVPMs and then
propose our quantum generalization. In the classical setting, there
are several proposals for ``imprecise
probabilities''~\cite{Dempster1967,Shafer1976,GilboaSchmeidler1994,Marinacci1999,Weichselberger2000,JamisonLodwick2004,HuberRonchetti2009,Grabisch2016}.
Although these proposals differ in some details, they all share the
fact that the probability $\mu(E)$ of an event~$E$ is generalized from
a single \emph{real number} to an \emph{interval}~$[\ell,r]$,
where~$\ell$ intuitively corresponds to the strength of evidence for
the event~$E$ and~$1-r$ corresponds to the strength of evidence
against the same event. Under some additional assumptions, this
interval could be interpreted as the Gaussian width of a probability
distribution.

We next introduce probability axioms for IVPMs. First, for each
interval $[\ell,r]$ we have the natural constraint
$0 \leq \ell \leq r \leq 1$ that guarantees that every element of the
interval can be interpreted as a conventional probability. We also
include $\imposs=[0,0]$ and $\necess=[1,1]$ as limiting intervals that
refer, respectively, to the probability interval for impossible events and
for events that are certain. We can write the latter as
$\mu(\emptyset)=\imposs$ and $\mu(\Omega)=\necess$, where $\emptyset$
is the empty set and $\Omega$ is the event covering the entire sample
space.  For each interval $[\ell,r]$, we also need the dual interval
$[1-r,1-\ell]$ so that if one interval refers to the probability of an
event~$E$, the dual refers to the probability of the event's
complement $\overline{E}$.  For example, if we discover as a result of
an experiment that $\mu(E) = [0.2,0.3]$ for some event~$E$, we may
conclude that $\mu\left(\overline{E}\right) = [0.7,0.8]$ for the
complementary event~$\overline{E}$. In addition to these simple
conditions, there are some subtle conditions on how intervals are
combined, which we discuss next.

Let $E_1$ and $E_2$ be two disjoint events with probabilities
$\mu(E_1)=[\ell_1,r_1]$ and $\mu(E_2)=[\ell_2,r_2]$. A first attempt
at calculating the probability of the combined event that
\emph{either} $E_1$ or $E_2$ occurs might be
$\mu(E_1\cup E_2) = [\ell_1+\ell_2,r_1+r_2]$. In some cases, this is
indeed a sensible definition. For example, if $\mu(E_1)=[0.1,0.2]$ and
$\mu(E_2)=[0.3,0.4]$ we get $\mu(E_1\cup E_2) = [0.4,0.6]$. But
consider an event $E$ such that $\mu(E)=[0.2,0.3]$ and hence
$\mu\left(\overline{E}\right)=[0.7,0.8]$. The two events $E$ and $\overline{E}$
are disjoint; the naïve addition of intervals would give
$\mu\left(E\cup\overline{E}\right)=[0.9,1.1]$, which is not a valid probability
interval. Moreover the event $E\cup\overline{E}$ is the
entire space; its probability interval should be
$\necess$ which is sharper than $[0.9,1.1]$. The problem is that the
two intervals are correlated: there is more information in the
combined event than in each event separately so the combined event should be mapped to
a sharper interval. In our example, even
though the ``true'' probability of $E$ can be anywhere in the range
$[0.2,0.3]$ and the ``true'' probability of~$\overline{E}$ can be
anywhere in the range $[0.7,0.8]$, the values are not independent. Any
value of $\mu(E) \leq 0.25$ will force $\mu\left(\overline{E}\right)\geq
0.75$. To account for such subtleties, the axioms of interval-valued
probability do not use a strict equality for the combination of
disjoint events. The correct constraint enforcing coherence of
the probability assignment for
$E_1\cup E_2$ when $E_1$ and $E_2$ are disjoint is taken to be:
\begin{equation}
\label{eq:disjoint}
\mu(E_1\cup E_2) \subseteq [\ell_1+\ell_2,r_1+r_2]\,.
\end{equation}
Note that for any event $E$ with $\mu(E)=[\ell,r]$, we always have
$\mu(\Omega)=\necess\subseteq[\ell,r]+[1-r,1-\ell]=\mu(E)\cup\mu\left(\overline{E}\right)$.

When combining non-disjoint events, there is a further subtlety whose
resolution will give us the final general condition for IVPMs. For
events $E_1$ and~$E_2$, not necessarily disjoint, we have:
\begin{equation}
\mu(E_1\cup E_2) + \mu(E_1\cap E_2) \subseteq \mu(E_1) + \mu(E_2)\,,
\label{eq:classicalconvex}
\end{equation}
which is a generalization of the classical inclusion-exclusion
principle that uses $\subseteq$ instead of $=$ for the same reason as
before. The new condition, known as
\emph{convexity}~\cite{Shapley1971,GilboaSchmeidler1994,NgMoYeh1997,Marinacci1999,MarinacciMontrucchio2005,Grabisch2016},
reduces to the previously motivated Eq.~(\ref{eq:disjoint}) when the
events are disjoint, i.e., when $\mu(E_1\cap E_2) = 0$.

We now have the necessary ingredients to define the quantum extension,
QIVPMs, as a generalization of both classical IVPMs and conventional
quantum probability measures. We will show that QIVPMs reduce to
classical IVPMs when the space of quantum events $\events$ is
restricted to mutually commuting events $\eventsC$, i.e.,
to compatible events that can be measured simultaneously. In
Sec.~\ref{sec:Gleason} we will discuss the connection between QIVPMs
and conventional quantum probability measures in detail.

\begin{definition}[QIVPM]\label{def:QIVPM}
  Assume a collection of intervals $\mathscr{I}$ including $\imposs$
  and $\necess$ with addition and scalar multiplication defined as
  follows:
  \begin{subequations}\label{eq:interval-operations}
  \begin{eqnarray}
   &  & [\ell_{1},r_{1}]+[\ell_{2},r_{2}]=[\ell_{1}+\ell_{2},r_{1}+r_{2}]\textrm{ and}\\
   &  & x[\ell,r]=\begin{cases}
  [x\ell,xr] & \textrm{for }x\ge0\,;\\{}
  [xr,x\ell] & \textrm{for }x\le0\,.
  \end{cases}\label{eq:interval-times}
  \end{eqnarray}
  \end{subequations}
  Then we take a QIVPM~$\bar{\mu}$ to be an assignment of an interval to each
  event (projection operator~$P$) subject to the following constraints:
  \begin{subequations}\label{eq:QIVPM-constraints}
  \begin{eqnarray}
   &  & \bar{\mu}(\mathbb{0})=\imposs\,,\\
   &  & \bar{\mu}(\mathbb{1})=\necess\,,\label{eq:necess}\\
   &  & \bar{\mu}\left(\mathbb{1}-P\right)= \necess-\bar{\mu}\left(P\right)\,,\label{eq:complement}
  \end{eqnarray}
  \end{subequations}
  and satisfying for each pair of \emph{commuting} projectors~$P_0$
  and~$P_1$ with $P_0P_1=P_1P_0$,
\begin{equation}
\bar{\mu}\left(P_{0}+P_{1}-P_{0}P_{1}\right)+\bar{\mu}\left(P_{0}P_{1}\right)\subseteq\bar{\mu}\left(P_{0}\right)+\bar{\mu}\left(P_{1}\right)\,.
\label{eq:QuantumInterval-valuedProbability-Inclusion}
\end{equation}
\end{definition}

\noindent The first three constraints,
Eqs.~(\ref{eq:QIVPM-constraints}), are the direct counterpart of the
corresponding ones for classical IVPMs.  Note that the minus sign
appearing in Eq.~(\ref{eq:complement}) is accommodated by the $x\le0$ case in
Eq.~(\ref{eq:interval-times}).  With the understanding that
the union of classical sets $E_1\cup E_2$ is replaced by
$P_0+P_1-P_0P_1$ in the case of quantum projection
operators~\cite{Griffiths2003}, the last condition,
Eq.~(\ref{eq:QuantumInterval-valuedProbability-Inclusion}), is a
direct counterpart of the convexity condition of
Eq.~(\ref{eq:classicalconvex}). Thus our definition of QIVPMs merges
aspects of both classical IVPMs and quantum probability measures.

Our definition of QIVPMs is consistent with classical IVPMs in the
sense that a restriction of QIVPMs to mutually commuting events,
$\eventsC$, recovers the definition of classical
IVPMs~\cite{JamisonLodwick2004}. The proof of this fact is included in the
forthcoming thesis by the first author~\cite{TaiThesis2018}. A
consequence is that known properties of classical IVPMs
directly hold for QIVPMs when one restricts to mutually commuting
events, $\eventsC$. In particular, in the classical world, it is
\emph{impossible} for experiments to result in probabilities that are
inconsistent with \emph{some} state of the system under consideration,
i.e., all IVPMs must have a non-empty ``core''~\footnote{A result by
  Shapley~\cite{Shapley1971,GilboaSchmeidler1994,NgMoYeh1997,Grabisch2016}
  proves that a classical IVPM always contains at least one state that
  is consistent with every event.}. Interestingly, as we show in
Sec.~\ref{sec:Gleason}, it is possible in the quantum world for the
probabilities associated with some events to be inconsistent with
\emph{any} quantum state, i.e., for the QIVPM to have an empty core;
in that case, one cannot guarantee non-empty cores for finite-precision
attempts at proving Gleason's theorem by extending the Born measure
$\muB_{\rho}\left(P\right)$ to QIVPMs
$\bar{\mu}\left(P\right)$. However, if we restrict ourselves to the set
$\eventsC$ of mutually commuting events, 
the situation reverts to the classical case in which
probabilities always determine at least one state.

We now give the necessary technical definitions to prove this
non-empty core property.
\begin{definition}[Consistency]\label{def:Consistency} We say a QIVPM
$\bar{\mu}$ is
\emph{consistent} with a state $\rho$ and a projector~$P$ if the
interval $\bar{\mu}(P)$ contains the exact probability calculated by
the Born rule~\cite{Born1983bibTeX,Mermin2007,Jaeger2007}, i.e.,
  \begin{equation}
  \muB_{\rho}\left(P\right) = \Tr\left(\rho
  P\right)\in\bar{\mu}\left(P\right)\ .\label{eq:consistent}
  \end{equation}
In contrast with classical probability spaces~\cite{Note1}, there is \emph{no
guarantee} that there exists a state $\rho$ that satisfies
Eq.~(\ref{eq:consistent}) and therefore is consistent with a QIVPM.
\end{definition}

We next refine the concept of consistency by introducing the idea of
a ``core'' set of states relative to subspaces. First, we define
$\events'$ as a \emph{subspace} of a set of events~$\events$ if
$\events'$ contains the projectors $\mathbb{0}$ and $\mathbb{1}$ and is closed under
complements, sums, and products.  In particular, for any
projector~$P\in\events'$, we have
$\mathbb{1}-P\in\events'$ and for each pair of commuting
projectors~$P_{0}\in\events'$ and $P_{1}\in\events'$, we have
$P_{0}+P_{1}-P_{0}P_{1}$ and $P_{0}P_{1}\in\events'$.

\begin{definition}[The core of a probability measure]\label{def:core} The
\emph{core}
$\coreBorn\left(\bar{\mu},\events'\right)$ of a probability measure 
$\bar{\mu}$ relative to a subspace of events $\events'$ is the
collection of all states $\rho$ that are \emph{consistent} with 
$\bar{\mu}$ on every projector in $\events'$, that is
\begin{equation}
\label{eq:hbar}
\coreBorn\left(\bar{\mu},\events'\right) = \set{\rho}{\forall P\in
\events',~\muB_{\rho}\left(P\right)\in\bar{\mu}\left(P\right)} \ .
\end{equation}
\end{definition}

We are now in position to state and prove that, for the special case
of commuting events, a QIVPM will always have a non-empty core.

\begin{thm}[Non-empty Core for Compatible Measurements] \label{thm:Shapley}
  For every QIVPM~$\bar{\mu}:\events\rightarrow\mathscr{I}$, if a subspace
  of events~$\eventsC\subseteq\events$ commutes, then
  $\coreBorn\left(\bar{\mu},\eventsC\right)\ne\emptyset$.
\end{thm}

An outline of the proof, detailed in the forthcoming
thesis~\cite{TaiThesis2018}, proceeds as
follows. From a subspace $\eventsC$ of mutually commuting events, one
can construct a partial orthonormal basis by diagonalization, and
complete this to a full orthonormal basis $\overline{\eventsC}$.  We
can then build a bijection between the QIVPM on the set of projectors
associated with this basis and the set of classical events
corresponding to this basis.  Using this correspondence together with
the classical result by
Shapley~\cite{Shapley1971,GilboaSchmeidler1994,NgMoYeh1997,Grabisch2016},
we can establish that for the special case of commuting events, a
QIVPM will always have a non-empty core.

We conclude this section with a generalization of expectation values
of observables in the context of QIVPMs. In conventional quantum
mechanics the expectation value of an observable as defined in
Eq.~\eqref{eq:quantum-expectation} is a unique real number. The
generalization to QIVPMs implies that this expectation value will
itself become bounded by an interval. 

\begin{definition}[Expectation Value of Observables over QIVPMs] Let
  $\mathscr{I}$ be a set of intervals; $\Hilb$ a Hilbert space of
  dimension $d$ with event space~$\events$; and $\mathbf{O}$ an
  observable with spectral decomposition
  $\sum_{i=1}^d \lambda_iP_i$. Let $\events'$ be the minimal
  subspace of events containing all the projectors $P_i$ in the
  spectral decomposition of $\mathbf{O}$ and define:
\begin{equation}
\expval{\mathbf{O}}_{\bar{\mu}}=\left[\min_{\rho\in\coreBorn\left(\bar{\mu},\events'\right)}
\expval{\mathbf{O}}_{\muB_{\rho}},\max_{\rho\in\coreBorn\left(\bar{\mu},\events'\right)}
\expval{\mathbf{O}}_{\muB_{\rho}}\right]\,. \label{eq:quantum-interval-expectation-core}
\end{equation}
\end{definition}

\noindent Intuitively the expectation value of an observable relative
to a QIVPM $\bar{\mu}$ lies between two possible outcomes, which
themselves lie between the minimum and maximum bounds of the
probability intervals associated with each state~$\rho$ that is
consistent with~$\bar{\mu}$ on every projector in the spectral
decomposition of the observable. If $\bar{\mu}$ is a conventional
(Born) probability measure induced by a state $\rho$, then the Born
rule probability induced by every state in
$\coreBorn\left(\bar{\mu},\events'\right)$ will be
$\muB_{\rho}$ and the interval collapses to a point, thus
reducing the definition to that of
Eq.~(\ref{eq:quantum-expectation})~\cite{TaiThesis2018}.
We also note that, when restricted to commuting projectors,
Eq.~(\ref{eq:quantum-interval-expectation-core}) is consistent with
the classical notion of the \emph{Choquet
  integral}~\cite{Choquet1954,GilboaSchmeidler1994,Grabisch2016}
which is used to calculate the expectation value of random variables
as a weighted average~\cite{TaiThesis2018}.

\section{The Kochen-Specker Theorem and Contextuality}
\label{sec:Kochen-Specker}
  
Our generalization of quantum probability measures to QIVPMs allows us
to strengthen the scope of one of the fundamental theorems of quantum
physics: the Kochen-Specker
theorem~\cite{BELL_1966,kochenspecker1967,Redhead1987-REDINA,Mermin1990Simple,peres1995quantum,Jaeger2007,Held2016}. Our
finite-precision extension of that theorem will suggest a resolution
to the debate initiated by Meyer and Mermin on the relevance of the
Kochen-Specker to experimental, and hence finite-precision, quantum
measurements~\cite{PhysRevLett.83.3751,HKSS1999apsrev4,Mermin1999,Kent1999,SimonBruknerZeilinger2001,Cabello2002,Larsson2002,Appleby2002,BarrettKent2004,Appleby_2005,Spekkens2005,GuehneKleinmannCabelloEtAl2010,MazurekPuseyKunjwalEtAl2016}.
Specifically, the original Kochen-Specker theorem is formulated using
a model quantum mechanical system that has \emph{definite values at
  all times}~\cite{Held2016}, i.e., its observables have infinitely
precise values at all times. Our interval-valued probability framework
will allow us to state, and prove, a stronger version of the theorem
that holds even if the observables have values that are only definite
up to some precision specified by a parameter $\delta$. Our approach
provides a quantitative realization of Mermin's
intuition~\cite{Mermin1999}:
\begin{quote}
  \ldots although the outcomes deduced from such imperfect
  measurements will occasionally differ dramatically from those
  allowed in the ideal case, if the misalignment is very slight, the
  statistical distribution of outcomes will differ only slightly from
  the ideal case.
\end{quote}


\subsection{Finite-Precision Extension of the Kochen-Specker Theorem}

The first step in our formalization is to introduce a family of QIVPMs
parameterized by an uncertainty~$\delta$, which we call
$\delta$-deterministic QIVPMs.

\begin{definition}[$\delta$-Determinism]\label{def:delta-deterministic} A
  QIVPM~$\bar{\mu}:\events\rightarrow\mathscr{I}$ is
  $\delta$-deterministic if, for every event $P\in\events$, we have
  that either 
  $\bar{\mu}\left(P\right)\subseteq\left[0,\delta\right]$ or
  $\bar{\mu}\left(P\right)\subseteq\left[1-\delta,1\right]$. 
\end{definition}


\noindent This definition puts no restrictions on the set of intervals
itself, only on which intervals are assigned to events. When
$\delta=0$, every event must be assigned a probability either in
$\imposs$ or in $\necess$, i.e., every event is completely determined
with certainty. As $\delta$ gets larger, the QIVPM allows for more
indeterminate behavior. 

The expectation value of an observable $\mathbf{O}$ in a Hilbert space
${\cal H}$ of dimension $d$ relative to a 0-deterministic QIVPM is
fully determinate and is equal to one of the eigenvalues $\lambda_i$
of that observable. To see this, note that given an orthonormal basis
$\Omega=\{ \ket{\psi_0},\ket{\psi_1},\ldots,\ket{\psi_{d-1}}\}$, a
0-deterministic QIVPM must map exactly one of the projectors
$\proj{\psi_i}$ to $\necess$ and all others to $\imposs$. This is
because, by Eq.~(\ref{eq:necess}), we have
$\bar{\mu}\left(\sum_{j=0}^{d-1}{\proj{\psi_j}}\right)=\necess$ and by
inductively applying
Eq.~(\ref{eq:QuantumInterval-valuedProbability-Inclusion}), we must
have one of the $\bar{\mu}(\proj{\psi_i})=\necess$ and all others
mapped to $\imposs$. Given any state~$\rho$ that is consistent with
this QIVPM on all the projectors in $\Omega$, we have by
Eq.~(\ref{eq:hbar}) that $\muB_{\rho}$ must also map exactly one of
the projectors in $\Omega$ to 1 and all others to 0. If an observable
has a spectral decomposition along~$\Omega$ then, by
Eq.~(\ref{eq:quantum-expectation}), its expectation value relative
to~$\muB_{\rho}$ is the eigenvalue $\lambda_i$ whose projector is
mapped to 1. It therefore follows, by
Eq.~(\ref{eq:quantum-interval-expectation-core}), that the expectation
value relative to the 0-deterministic~${\bar{\mu}}$ is fully
determinate and lies in the interval $[\lambda_i,\lambda_i]$.

We can now proceed with the main technical result of this section. We
first observe that the original Kochen-Specker theorem is a statement
regarding the non-existence of a $0$-deterministic QIVPM, and
generalize to a corresponding statement about $\delta$-deterministic QIVPMs.

\begin{thm}[0-Deterministic Variant of the Kochen-Specker
  Theorem] \label{thm:Kochen-Specker} Given a Hilbert space $\Hilb$ of
  dimension $d\ge3$, there is no $0$-deterministic measure~$\bar{\mu}$
  mapping every event to either $\imposs$ or $\necess$.
\end{thm}

\noindent To explain why this result is equivalent to the original
Kochen-Specker theorem and to prove it at the same time, we proceed by
assuming a 0-deterministic QIVPM $\bar{\mu}$ and derive the same
contradiction as the original Kochen-Specker theorem. Instead of
adapting the more complicated proof for $d=3$, the counterexample
presented below uses the simpler proof for a Hilbert space of
dimension $d=4$ and is constructed as follows.

We consider a two spin-$\frac{1}{2}$ Hilbert space
$\Hilb=\Hilb_1\otimes\Hilb_2$ of dimension $d=4$. We use the same nine
observables $\mathbf{O}_{ij}$ with $i$ and $j$ ranging over
$\{0,1,2\}$ from the Mermin-Peres ``magic square'' used to prove the
Kochen-Specker
theorem~\cite{Mermin1990Simple,peres1995quantum,Griffiths2003}:

{\renewcommand{\arraystretch}{2}%
\begin{center}
\begin{tabular}{r|@{\quad}c@{\quad}|@{\quad}c@{\quad}|@{\quad}c@{\quad}|}
$\mathbf{O}_{ij}$~ & $j=0$ & $j=1$ & $j=2$ \\
\hline 
$i=0~$ & $\mathbb{1}\otimes\sigma_{z}$  & $\sigma_{z}\otimes\mathbb{1}$  & $\sigma_{z}\otimes\sigma_{z}$ \tabularnewline
\hline 
$i=1~$ & $\sigma_{x}\otimes\mathbb{1}$  & $\mathbb{1}\otimes\sigma_{x}$  & $\sigma_{x}\otimes\sigma_{x}$ \tabularnewline
\hline 
$i=2~$ & $\sigma_{x}\otimes\sigma_{z}$  & $\sigma_{z}\otimes\sigma_{x}$  & $\sigma_{y}\otimes\sigma_{y}$ \tabularnewline
\hline 
\end{tabular}
\par\end{center}
} 

\noindent The observables are constructed using the Pauli matrices
$\left\{ \mathbb{1},\sigma_{x},\sigma_{y},\sigma_{z}\right\}$ whose
eigenvalues are all either~$1$ or $-1$
\cite{Redhead1987-REDINA,544199,Griffiths2003,Jaeger2007,Mermin2007}.
They are arranged such that in
each row and column, \emph{except the column $j=2$}, every observable
is the product of the other two. In the $j=2$ column, we have instead
that
$\left(\sigma_{z}\otimes\sigma_{z}\right)\left(\sigma_{x}\otimes\sigma_{x}\right)=-\sigma_{y}\otimes\sigma_{y}$. Now
assume a 0-deterministic QIVPM~${\bar{\mu}}$; the expectation values
of the observables in each row relative to this 0-deterministic QIVPM
are fully determinate and must lie in either the interval $[1,1]$ or
the interval $[-1,-1]$ depending on which eigenvalue is the one whose
associated projector is certain. Since the product of any two
observables in a row is equal to the third, there must be an even
number of occurrences of the interval $[-1,-1]$ in each row and hence
in the entire table~\cite{TaiThesis2018}. However, looking at the expectation values of the
observables in each column, there must be an even number of
occurrences of the interval $[-1,-1]$ in the first two columns and an
odd number in the $j=2$ column and hence in the entire table~\cite{TaiThesis2018}. The
contradiction implies the non-existence of the assumed 0-deterministic
QIVPM. 

Our framework allows us to generalize the above theorem to state that
for small enough $\delta$, it is impossible to have
$\delta$-deterministic QIVPMs. 

We next prove a main result of this paper which is a stronger
statement of contextuality that includes the effects of
finite-precision.  Every QIVPM must map some events to truly uncertain
intervals, not just ``almost definite intervals.'' The proof requires
two simple lemmas that we present first. 

The first lemma shows a simpler way to prove the convexity
condition. Recall that the convexity condition for a QIVPM
$\bar{\mu} : \events \rightarrow \mathscr{I}$ states that for each pair of
\emph{commuting} projectors~$P$ and~$P'$ with $PP'=P'P$, the following
equation holds:
\begin{equation}
\bar{\mu}\left(P+P'-PP'\right)+\bar{\mu}\left(PP'\right)\subseteq\bar{\mu}\left(P\right)+\bar{\mu}\left(P'\right)\,.
\end{equation}

\begin{lemma}\label{lemma:verify-convexity}
  To verify the convexity condition of a QIVPM
  $\bar{\mu} : \events \rightarrow \mathscr{I}$, it is sufficient to check
  that:
\begin{equation}
\bar{\mu}\left(P_{0}+P_{1}\right)=\bar{\mu}\left(P_{0}\right)+\bar{\mu}\left(P_{1}\right)\label{amr}
\end{equation}
 for all orthogonal projectors $P_{0}$ and~$P_{1}$.
\end{lemma}
\noindent The proof follows the outline of the proof of the classical
inclusion-exclusion principle. From the commuting projectors $P$ and
$P'$, we construct the following three orthogonal projectors: $PP'$,
$P(\mathbb{1}-P')$, and $(\mathbb{1}-P)P'$. Then we proceed as
follows:
\begin{widetext}
\begin{equation*}
\begin{aligned}
\noalign{$\bar{\mu}\left(P+P'-PP'\right)+\bar{\mu}\left(PP'\right)$}
\qquad\qquad=~~&\bar{\mu}\left(PP'+P(\mathbb{1}-P')+P'-PP'\right)+\bar{\mu}\left(PP'\right) 
  & (\mbox{because } P = PP' + P - PP') \\
=~~& \bar{\mu}\left(P(\mathbb{1}-P')+P'\right)+\bar{\mu}\left(PP'\right) \\
=~~& \bar{\mu}\left(P(\mathbb{1}-P')+PP'+(\mathbb{1}-P)P'\right)+\bar{\mu}\left(PP'\right) 
      & (\mbox{because } P' = PP' + P' - PP') \\
=~~& \bar{\mu}(P(\mathbb{1}-P'))+\bar{\mu}(PP')+\bar{\mu}((\mathbb{1}-P)P')+
    \bar{\mu}\left(PP'\right) & (\mbox{using Eq.~(\ref{amr}) twice}) \\
=~~& \bar{\mu}(P(\mathbb{1}-P')+PP')+\bar{\mu}((\mathbb{1}-P)P'+PP') & (\mbox{using Eq.~(\ref{amr}) twice}) \\
=~~& \bar{\mu}(P) + \bar{\mu}(P')
\end{aligned}
\end{equation*}
\end{widetext}

The next lemma relates $\delta$-deterministic QIVPMs with $\delta <
\frac{1}{3}$ to 0-deterministic QIVPMs. 

\begin{lemma}\label{lemma:delta2zero} 
  From any $\delta$-deterministic
  QIVPM~$\bar{\mu}:\events\rightarrow\mathscr{I}$ with
  $\delta < \frac{1}{3}$, we can construct a $0$-deterministic QIVPM
  $\bar{\mu}^{D}:\events\rightarrow\left\{
    \imposs,\necess\right\}$ defined as follows:
\begin{equation}
\bar{\mu}^{D}\left(P\right)=\begin{cases}
\imposs & \textrm{ if }\bar{\mu}\left(P\right)\subseteq\left[0,\delta\right]\:;\\
\necess & \textrm{ if }\bar{\mu}\left(P\right)\subseteq\left[1-\delta,1\right]\:.
\end{cases}
\end{equation}
\end{lemma}
\noindent The most important part of the proof is to verify the convexity
condition for $\bar{\mu}^{D}$. By Lemma~\ref{lemma:verify-convexity}, it is sufficient
to verify the following equation for orthogonal projectors $P_{0}$
and~$P_{1}$,
\begin{equation}
\bar{\mu}^{D}\left(P_{0}+P_{1}\right)=\bar{\mu}^{D}\left(P_{0}\right)+\bar{\mu}^{D}\left(P_{1}\right)\,, \label{eq:QuantumInterval-valuedProbability-Equal}
\end{equation}
for two conditions, which we now examine in detail.

When one of $\bar{\mu}^{D}\left(P_{0}\right)$ and $\bar{\mu}^{D}\left(P_{1}\right)$
is $\necess$, say $\bar{\mu}^{D}\left(P_{0}\right)=\imposs$
and $\bar{\mu}^{D}\left(P_{1}\right)=\necess$, we have $\bar{\mu}\left(P_{0}\right)\subseteq\left[0,\delta\right]$
and $\bar{\mu}\left(P_{1}\right)\subseteq\left[1-\delta,1\right]$ which
implies $\bar{\mu}\left(P_{0}+P_{1}\right)\subseteq\left[1-\delta,1+\delta\right]$.
Since $\bar{\mu}\left(P_{0}+P_{1}\right)$ is a subset of $\left[0,1\right]$,
$\bar{\mu}\left(P_{0}+P_{1}\right)$ must be a subset of $\left[1-\delta,1\right]$,
which implies $\bar{\mu}^{D}\left(P_{0}+P_{1}\right)$ is also
$\necess$, thus satisfying Eq.~(\ref{eq:QuantumInterval-valuedProbability-Equal}).

When both
$\bar{\mu}^{D}\left(P_{0}\right)$ and $\bar{\mu}^{D}\left(P_{1}\right)$
are $\imposs$, we have both $\bar{\mu}\left(P_{0}\right)$ and
$\bar{\mu}\left(P_{1}\right)\subseteq\left[0,\delta\right]$ which
implies
$\bar{\mu}\left(P_{0}+P_{1}\right)\subseteq\left[0,2\delta\right]$.
Since we assume $\delta<\frac{1}{3}$, $\left[0,2\delta\right]$ and
$\left[1-\delta,1\right]$ are disjoint, which implies
$\bar{\mu}\left(P_{0}+P_{1}\right)$ and $\left[1-\delta,1\right]$ are
disjoint. Together with the fact that
$\bar{\mu}\left(P_{0}+P_{1}\right)$ is a subset of either
$\left[0,\delta\right]$ or $\left[1-\delta,1\right]$,
$\bar{\mu}\left(P_{0}+P_{1}\right)$ must be a subset of
$\left[0,\delta\right]$, which implies
$\bar{\mu}^{D}\left(P_{0}+P_{1}\right)=\imposs$, and hence
also Eq.~(\ref{eq:QuantumInterval-valuedProbability-Equal}) is again satisfied.

\begin{thm}[Finite-precision Extension of the Kochen-Specker Theorem]
\label{cor:Kochen-Specker-IVPM} Given a Hilbert space $\Hilb$ of 
dimension~$d\ge3$, there is no $\delta$-deterministic QIVPM for 
$\delta<\frac{1}{3}$.\end{thm}
\noindent The proof is by contradiction: Suppose there is a
$\delta$-deterministic
QIVPM~$\bar{\mu}:\events\rightarrow\mathscr{I}$. By
Lemma~\ref{lemma:delta2zero}, we can construct a 0-deterministic QIVPM; however, by
Thm.~\ref{thm:Kochen-Specker}, such 0-deterministic QIVPMs do not
exist.

The bound $\delta < \frac{1}{3}$ is tight as it is possible to
construct a $\frac{1}{3}$-deterministic QIVPM
$\bar{\mu}:\events\rightarrow\mathscr{I}$. For example, consider a
three-dimensional Hilbert space~$\Hilb$ with orthonormal basis
$\left\{\ket{0},\ket{1},\ket{2}\right\}$ and a state
$\rho=\frac{1}{3}\proj{0}+\frac{1}{3}\proj{1}+\frac{1}{3}\proj{2}$.
Let $P$ be an operator projecting onto an $n$-dimensional subspace
of~$\Hilb$, where $n \leq 3$. It is straightforward to check that
$\muB_{\rho}\left(P\right)=\frac{n}{3}$. Therefore,
$\bar{\mu}\left(P\right)=\left[\muB_{\rho}\left(P\right),\muB_{\rho}\left(P\right)\right]$
is a valid $\frac{1}{3}$-deterministic QIVPM.

When $\delta \geq \frac{1}{3}$, i.e, when the uncertainty in
measurements becomes so large, it becomes possible to map every
observable to some (quite inaccurate) probability interval, thus
invalidating the Kochen-Specker theorem.

\begin{figure}
\begin{center}
\includegraphics[scale=0.5]{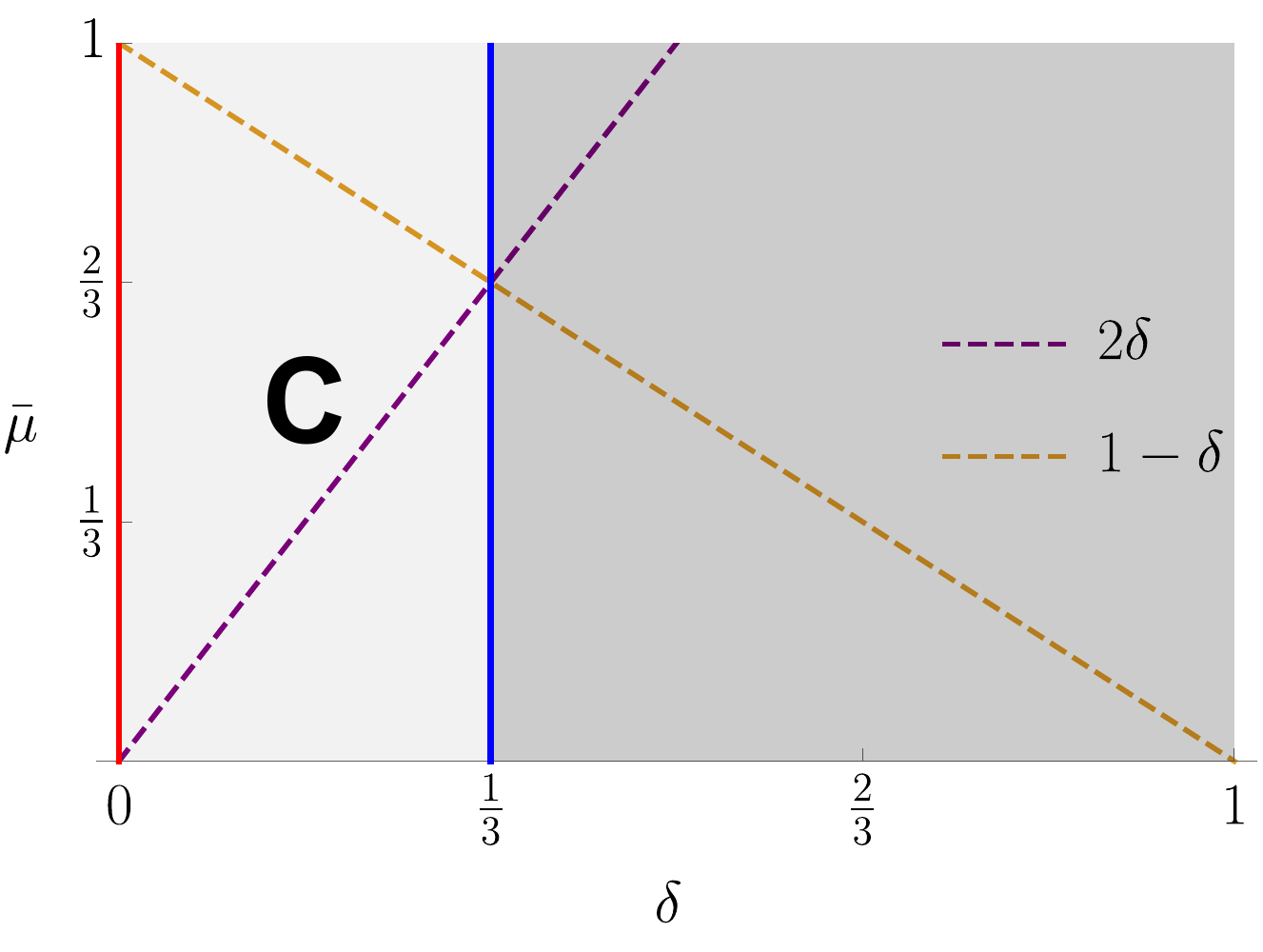} 
\par\end{center}
\caption{The region to the left of the vertical line at
$\delta=\frac{1}{3}$ is where we assume small measurement degradation;
in that region our extension of the KS theorem definitely demonstrates
contextuality ({\bf{\sf C}}). In the region to the right, the degradation of the data
is large and our extension of the KS theorem no longer refutes other
explanations for the experimental data.}
\label{fignoname}
\end{figure}

We can summarize and illustrate the above arguments using
Fig.~\ref{fignoname}. 

As is the case for conventional, infinitely-precise, quantum
probability measures, the theorem is only applicable to dimensions
$d \geq 3$. Indeed when the Hilbert space has dimension 2, it is
straightforward to construct a 0-deterministic QIVPM as
follows. Consider a non-contextual hidden variable model for $d=2$
(e.g., as proposed by Bell or
Kochen-Specker~\cite{BELL_1966,kochenspecker1967}). Such a
two-dimensional model assigns definite values to all observables at
all times, and hence assigns a \emph{determinate} probability (0 or 1)
to each event. This probability measure directly induces a
0-deterministic QIVPM by changing 0 to $\imposs$ and 1 to
$\necess$.  It follows that every 0-deterministic QIVPM is
$\delta$-deterministic.



\subsection{Experimental Data and \texorpdfstring{$\delta$}{delta}-determinism} 

\begin{table*}
\caption{\label{tab:probability-measures}Possible probability measures on
a Hilbert space of dimension~$d=3$, where $\bar{\mu}_{2}'$ and
$\bar{\mu}_{3}$ are QIVPMs while $\bar{\mu}_{0}$, $\bar{\mu}_{1}$,
and $\bar{\mu}_{2}$ are not. Events are listed in the column labeled
by $P$.}
\centering{}%
\begin{tabular}{cccccc}
\toprule 
\addlinespace
$P$  & $\bar{\mu}_{0}\left(P\right)$ & $\bar{\mu}_{1}\left(P\right)$ & $\bar{\mu}_{2}\left(P\right)$ & $\bar{\mu}_{2}'\left(P\right)$ & $\bar{\mu}_{3}\left(P\right)$\tabularnewline
\midrule
\midrule 
\addlinespace
$\mathbb{0}$  & $\imposs$ & $\imposs$ & $\imposs$ & $\imposs$ & $\imposs$\tabularnewline
\midrule 
\addlinespace
All one-dimensional projectors  & $\left[0,0\right]$ & $\left[0,\tfrac{1}{4}\right]$ & $\left[0,\tfrac{1}{3}\right]$ & $\left[\tfrac{1}{3},\tfrac{1}{3}\right]$ & $\left[0,\tfrac{1}{2}\right]$\tabularnewline
\midrule 
\addlinespace
All two-dimensional projectors  & $\left[1,1\right]$ & $\left[\tfrac{3}{4},1\right]$ & $\left[\tfrac{2}{3},1\right]$ & $\left[\tfrac{2}{3},\tfrac{2}{3}\right]$ & $\left[\tfrac{1}{2},1\right]$\tabularnewline
\midrule 
\addlinespace
$\mathbb{1}$  & $\necess$ & $\necess$ & $\necess$ & $\necess$ & $\necess$\tabularnewline
\bottomrule
\end{tabular}
\end{table*}

We have thus quantified one important aspect of uncertainty in quantum
mechanics---the effect of the imprecise nature of devices---which is a novel
addition to the theory of measurement. Indeed, as Heisenberg
emphasized in his famous microscope
example~\cite{Heisenberg1983apsrev4}, the conventional theory
of measurement states that it is impossible to precisely measure any
property of a system without disturbing it somewhat. Thus, there are
fundamental limits to what one can measure and these limits have traditionally 
been attributed to complementarity. Our imprecision represents an 
\emph{additional} source of indeterminacy beyond the inherent probabilistic nature
of quantum mechanics.




In an experimental setup, $\delta$ is calculated as follows. To
determine the probability of any event, we typically repeat an
experiment $m$ times and count the number of times we witness the
event. This assumes that for each run of the experiment we can
determine, using our apparatus, whether the event occurred or
not. Assume an event has an ideal mathematical probability of $0$, and
we repeat the experiment $100$ times. In a perfect world we should be
able to refute the event $100$ times and calculate that the
probability is $0$. We might also observe the event $2$ times and
refute it $98$ times and therefore calculate the probability to be
$0.02$. Note that this situation assumes perfect measurement
conditions and remains within the context of conventional
(real-valued) probability theory. The question we focus on is what
happens if we are only able to refute it $97$ times and are
\emph{uncertain} $3$ times? This is quite common in actual
experiments. Mathematically we can model this idea by stating that the
probability of the event is in the range $\left[0,0.03\right]$ which
says that the probability of the event could be $0$, $0.01$, $0.02$,
or $0.03$ as each the three uncertain records could either be evidence
for the event or against it. We just cannot nail it down given the
current experimental results and therefore represent the evidence as a
($\delta=$)$0.03$-deterministic probability measure. The interesting
observation is that the axioms of probability theory (like additivity
and convexity) impose enough constraints on the structure of
interval-valued quantum probability measures to make them robust in
the face of small non-vanishing $\delta$'s.

To see this idea in the context of a quantum experiment, consider
a three-dimensional Hilbert space with one-dimensional projectors~$P_{\rho}$,
two-dimensional projectors $P_{\rho}+P_{\sigma}$, and an experiment
that is repeated $12$ times. By the Kochen-Specker theorem, it is impossible
to build a probability measure that maps every projection to either
$0=\frac{0}{12}$ or $1=\frac{12}{12}$. That is, the assignment~$\bar{\mu}_{0}$
defined in Table~\ref{tab:probability-measures} is not a QIVPM.

Now consider what happens if $\frac{1}{4}$ of the data for \emph{every}
one-dimensional projector is uncertain. A potential account of this
degradation is to assign to each event $P$ the entire range of possibilities
$\bar{\mu}_{1}(P)$ as defined in Table~\ref{tab:probability-measures}.
This measure is not a valid QIVPM because it does not satisfy
the convexity condition: for any two orthogonal one-dimensional events
$P_{0}$ and $P_{1}$, the convexity condition requires $\bar{\mu}_{1}\left(P_{0}+P_{1}\right)\subseteq\bar{\mu}_{1}\left(P_{0}\right)+\bar{\mu}_{1}\left(P_{1}\right)$,
but $\bar{\mu}_{1}\left(P_{0}+P_{1}\right)=\left[\tfrac{3}{4},1\right]$
which is not a subset of $\left[0,\tfrac{1}{2}\right]=\bar{\mu}_{1}\left(P_{0}\right)+\bar{\mu}_{1}\left(P_{1}\right)$.
Interestingly, it is impossible to find any probability measure that
would be consistent with these observations, as the interval $\left[\tfrac{3}{4},1\right]$
is completely disjoint from the interval $\left[0,\tfrac{1}{2}\right]$
and no amount of shifting of assumptions regarding the precise outcome
of the uncertain observations could change that disjointness. However,
as shown next, a sharp transition occurs when $\delta=\tfrac{1}{3}$.

When the proportion of uncertain data reaches $\frac{1}{3}$, the probability measure
that assigns to each event the entire range of possibilities is
$\bar{\mu}_{2}$ defined in Table~\ref{tab:probability-measures}.
This is also not a valid probability measure by the same
argument as above. However, in this case $\bar{\mu}_{2}\left(P_{0}+P_{1}\right)=\left[\tfrac{2}{3},1\right]$
and $\left[0,\tfrac{2}{3}\right]=\bar{\mu}_{2}\left(P_{0}\right)+\bar{\mu}_{2}\left(P_{1}\right)$
have a \emph{common point}. Hence, by assuming that the uncertain data for
one-dimensional projectors always support the associated event, while
those for two-dimensional projectors always refute the event, we can
find the probability measure~$\bar{\mu}_{2}'$ that can verified to be a
valid QIVPM and is consistent with the experimental data.

A similar situation happens when more than $\frac{1}{3}$ of data
is uncertain. In particular, if half of the data is uncertain, the probability
measure~$\bar{\mu}_{3}$ that assigns to each event the entire range of possibilities
is already a QIVPM.

\section{The Born Rule and Gleason's Theorem}

\label{sec:Gleason}

A conventional quantum probability measure can be easily constructed
from a state $\rho$ according to the Born
rule~\cite{Born1983bibTeX,Mermin2007,Jaeger2007}.  According
to Gleason's
theorem~\cite{gleason1957,Redhead1987-REDINA,peres1995quantum}, this
state $\rho$ is also the unique state consistent with any possible
probability measure. 

\subsection{Finite-Precision Extension of Gleason's Theorem}

In order to re-examine these results in our
framework, we first reformulate Gleason's theorem in QIVPMs using
infinitely precise uncountable
intervals~$\mathscr{I}_{\infty}=\set{\left[x,x\right]}{x\in\left[0,1\right]}$:

\begin{thm}[$\mathscr{I}_{\infty}$ Variant of the Gleason
  Theorem]\label{cor:Gleason's}In
  a Hilbert space $\Hilb$ of dimension $d\geq3$, given a
  QIVPM~$\bar{\mu}:\events\rightarrow\mathscr{I}_{\infty}$, the state
  $\rho$ consistent with~$\bar{\mu}$ on every projector
  is unique, i.e., there exists a
  unique state~$\rho$ such that
  $\coreBorn\left(\bar{\mu},\events\right)=\{\rho\}$.
  \end{thm}

Now let us consider relaxing $\mathscr{I}$
to a countable set of finite-width intervals.
As the intervals in the image of a QIVPM become less and less sharp,
we expect more and more states to be consistent with it. In
the limit of minimal sharpness, all states~$\rho$
are consistent with the QIVPM
\begin{equation}
\bar{\mu}\left(P\right)=\begin{cases}
\imposs & \textrm{ if }P=\mathbb{0}\,;\\
\necess & \textrm{ if }P=\mathbb{1}\,;\\
\unknown=\left[0,1\right] & \textrm{ otherwise}
\end{cases}
\end{equation}%
mapping nearly all
projections to the \emph{unknown} interval~\unknown. There is however a
subtlety: as shown in
the theorem below, it is possible for an arbitrary assignment of intervals to projectors
to be globally inconsistent.


\begin{thm}[Empty Cores Exist for General
  QIVPMs]\label{thm:Non-extensible-of-Gleason's}There exists a Hilbert
  space  ${\cal H}$ and a QIVPM~$\bar{\mu}:\events\rightarrow\mathscr{I}$ such
  that $\coreBorn\left(\bar{\mu},\events\right)=\emptyset$.\end{thm}

To prove this theorem, we need to construct a QIVPM on some Hilbert
space, and verify that there are no states that are consistent (see
Defs.~\ref{def:Consistency} and \ref{def:core}) with it
on all possible events. Assume a Hilbert space of dimension $d=3$ with
orthonormal basis $\left\{ \ket{0},\ket{1},\ket{2}\right\} $, let
$\ket{\ps}=\left(\ket{0}+\ket{1}\right)/\sqrt{2}$,
$\ket{\ps'}=\left(\ket{0}+\ket{2}\right)/\sqrt{2}$, and assign
\begin{equation}
\mathscr{I}_{0}=\left\{ \necess,\imposs,\unknown\right\} \,.\label{eq:3-value-intervals}
\end{equation}
The map~$\bar{\mu}:\events\rightarrow\mathscr{I}_{0}$ defined in
Table~\ref{tab:non-Born-QIVPM} can be verified to be a
QIVPM~\cite{TaiThesis2018}.
\begin{table}
\caption{\label{tab:non-Born-QIVPM}QIVPM~$\bar{\mu}:\events\rightarrow\mathscr{I}_{0}$
on a Hilbert space of dimension~$d=3$. Events are listed in the column
labeled by $P$.}

\begin{tabular}{cc}
\toprule 
\addlinespace
$P$ & $\bar{\mu}\left(P\right)$\tabularnewline\addlinespace
\midrule
\midrule 
\addlinespace
$\mathbb{0}$, $\proj{0}$, $\proj{\ps}$, $\proj{\ps'}$ & $\imposs$\tabularnewline\addlinespace
\midrule 
\addlinespace
$\mathbb{1}$, $\mathbb{1}-\proj{0}$, $\mathbb{1}-\proj{\ps}$, 
$\mathbb{1}-\proj{\ps'}$ & $\necess$\tabularnewline\addlinespace
\midrule 
\addlinespace
All other projectors & $\unknown$\tabularnewline\addlinespace
\bottomrule
\end{tabular}
\end{table}
Next we will prove by contradiction that
$\coreBorn\left(\bar{\mu},\events\right)$ is the empty set. Suppose
there is a state
$\rho=\sum_{j=1}^{N}q_{j}\proj{\phi_{j}}\in\coreBorn\left(\bar{\mu},\events\right)$,
where $\sum_{j=1}^{N}q_{j}=1$ and $q_{j} > 0$. Since
we assumed the core $\coreBorn\left(\bar{\mu},\events\right)$ 
is non-empty, so  $\muB_{\rho}(P)\in\bar{\mu}(P)$, and 
Table~\ref{tab:non-Born-QIVPM} tells us that
$\bar{\mu}(\proj{0})=\imposs=[0,0]$, we must conclude that
$\muB_{\rho}(\proj{0})=0\in[0,0]$, and similarly for $\proj{\ps}$ and $\proj{\ps'}$.
If this is true, then $\ip{0}{\phi_{j}}=\ip{\ps}{\phi_{j}}=\ip{\ps'}{\phi_{j}}=0$
for all~$j$, and thus
\begin{subequations}
\begin{eqnarray}
 &  & \ip{1}{\phi_{j}}=\sqrt{2}\ip{\ps}{\phi_{j}}-\ip{0}{\phi_{j}}=0\,,\\
 &  & \ip{2}{\phi_{j}}=\sqrt{2}\ip{\ps'}{\phi_{j}}-\ip{0}{\phi_{j}}=0\,.
\end{eqnarray}
\end{subequations}
The above equations imply
$\ket{\phi_{j}}=\ket{0}\ip{0}{\phi_{j}}+\ket{1}\ip{1}{\phi_{j}}+\ket{2}\ip{2}{\phi_{j}}=0$,
violating the assumption that $\ket{\phi_{j}}$ is a normalized state, and thus the
theorem is proved.

The fact that a collection of poor measurements on a quantum system
cannot reveal the underlying state is not surprising. Under certain
conditions, we can however guarantee that the uncertainty in measurements
is consistent with \emph{some} non-empty collection of quantum states.
Furthermore, we can relate the uncertainty in measurements to the
volume of quantum states such that, in the limit of infinitely precise
measurements, the volume of states collapses to a single state.

To that end, we introduce the concept of \emph{interval maps}, which
we can use to construct a consistent family of QIVPMs.
An interval map $f:\left[0,1\right]\rightarrow\mathscr{I}$
maps every real-valued probability $x\in\left[0,1\right]$ to a set
of intervals $f\left(x\right)=\left[\ell,r\right]$ containing $x$,
where $\left[0,1\right]$ denotes the set of real-valued probabilities
(this should not be confused with
the interval-valued probability $\unknown$).
We also need a notion of \emph{norm} to quantify the
distance between (pure or mixed) states. The norm of a pure state
$\rho=\proj{\psi}$ is defined as usual by $\left\Vert \psi\right\Vert =\sqrt{\ip{\psi}{\psi}}$.
For any given Hermitian operator~$A$, we choose the
operator norm $\left\Vert A\right\Vert =\max_{\left\Vert \psi\right\Vert =1}\left\Vert A\ket{\psi}\right\Vert $,
which is also known as the $2$-norm or the spectral norm~\cite{RobertsVarberg1973,peres1995quantum,GolubVanLoan1996,Foucart2012}.
In fact, for any such matrix, including the density matrix~$\rho$,
this norm is the eigenvalue with maximum absolute value. Then, a
finite-precision extension of Gleason's theorem can be stated as
follows:

\begin{thm}[Finite-Precision Extension of the Gleason Theorem]\label{thm:Finite-precision-Gleason}Let
$f:\left[0,1\right]\rightarrow\mathscr{I}$ be an interval map
and let the composition $f\circ\muB_{\rho}$ be a QIVPM, where
$\muB_{\rho}$ is the
probability measure induced by the Born rule for a given state~$\rho$.
Let $\alpha$ be the maximum length of intervals in $\mathscr{I}$.
If a state $\rho'$ is consistent with $f\circ\muB_{\rho}$ on all
events, i.e., $\rho'\in\coreBorn\left(f\circ\muB_{\rho},\events\right)$,
then the norm of their difference is bounded by $\alpha$, i.e., $\left\Vert \rho-\rho'\right\Vert \le\alpha$.\end{thm}

The proof proceeds as follows. Given a state~$\rho'$ consistent
with $f\circ\muB_{\rho}$, we have $\muB_{\rho'}\left(\proj{\psi}\right)\in f\left(\muB_{\rho}\left(\proj{\psi}\right)\right)$
for any one-dimensional projector $P=\proj{\psi}$. Since the maximum
length of the intervals in $\mathscr{I}$ is $\alpha$, it is also
the upper bound of the difference: 
\[
\left|\muB_{\rho'}\left(\proj{\psi}\right)-\muB_{\rho}\left(\proj{\psi}\right)\right|=\left|\melem{\psi}{\rho-\rho'}{\psi}\right|\le\alpha\,.
\]
Since $\rho-\rho'$ is Hermitian, $\max_{\left\Vert \psi\right\Vert =1}\left|\melem{\psi}{\rho-\rho'}{\psi}\right|$
is the maximum absolute value of the eigenvalues of $\rho-\rho'$~\cite{544199},
and equal to $\left\Vert \rho-\rho'\right\Vert $~\cite{GolubVanLoan1996,Foucart2012}.
Hence, $\left\Vert \rho-\rho'\right\Vert \le\alpha$.

\subsection{Ultramodular Functions}

Theorem~\ref{thm:Finite-precision-Gleason} generalizes Gleason's
theorem in the sense that it accounts for a larger class of probability measures that includes
the conventional one as a limit. The theorem is however ``special'' in
the sense that it only applies to the particular class of QIVPMs
constructed by composing an interval map with a conventional
quantum probability measure. QIVPMs constructed in this manner have
some peculiar properties that we examine next.

An interval map is called \emph{ultramodular} if it satisfies
the following properties:

\begin{definition}[Ultramodular Functions]\label{def:THOS}Given
  a collection of intervals $\mathscr{I}$ including $\imposs$ and
  $\necess$, an interval map
  $\ultramodular:\left[0,1\right]\rightarrow\mathscr{I}$ is called
  ultramodular if:
\begin{subequations}\label{eq:iota-constraints}
\begin{eqnarray}
 &  & \ultramodular(0)=\imposs\,,\\
 &  & \ultramodular(1)=\necess\,,\\
 &  & \ultramodular\left(1-x\right)=\necess-\ultramodular\left(x\right)\,,
\end{eqnarray}
\end{subequations}and for any three numbers~$x_{0}$, $x_{1}$, and
$x_{2}\in\left[0,1\right]$ such that
$y=x_{0}+x_{1}+x_{2}\in\left[0,1\right]$, we have:
\begin{equation}
\ultramodular\left(y\right)+\ultramodular\left(x_{2}\right)\subseteq\ultramodular\left(x_{0}+x_{2}\right)+\ultramodular\left(x_{1}+x_{2}\right)\,.\label{eq:iota-Inclusion}
\end{equation}
\end{definition}

\noindent The first three constraints,
Eqs.~(\ref{eq:iota-constraints}), are the direct counterpart of the
corresponding QIVPM constraints, Eqs.~(\ref{eq:QIVPM-constraints});
the last condition, Eq.~(\ref{eq:iota-Inclusion}), is the direct
counterpart of the convexity conditions,
Eqs.~(\ref{eq:classicalconvex}) and
(\ref{eq:QuantumInterval-valuedProbability-Inclusion})
\cite{Choquet1954,Shapley1971,NgMoYeh1997,MarinacciMontrucchio2005}. Therefore,
these conditions guarantee that, for any conventional quantum
probability measure $\mu$, the composition $\ultramodular \circ \mu$
defines a valid QIVPM. Conversely, if for every quantum probability
measure $\mu$, it is the case that $f \circ \mu$ is a QIVPM, then
the interval map~$f$ is an ultramodular function. Formally, we have the following
result:

\begin{thm}[Equivalence of Ultramodular Functions and IVPMs]\label{thm:iota-statements}The
following three statements are equivalent:
\begin{enumerate}
\item \label{enu:iota-subject-to}A function~$\ultramodular:\left[0,1\right]\rightarrow\mathscr{I}$
is ultramodular.
\item \label{enu:iota-mu-CIVPM}The composite
  function~$\ultramodular\circ\mu:\eventsC\rightarrow\mathscr{I}$
  is a classical IVPM for all classical probability
  measures~$\mu:\eventsC\rightarrow\left[0,1\right]$.
\item \label{enu:iota-mu-QIVPM}The composite
  function~$\ultramodular\circ\mu:\events\rightarrow\mathscr{I}$ is a
  QIVPM for all quantum probability
  measures~$\mu:\events\rightarrow\left[0,1\right]$.
\end{enumerate}
\end{thm}

Statement~\ref{enu:iota-subject-to} implies \ref{enu:iota-mu-CIVPM}
and~\ref{enu:iota-mu-QIVPM} as we have outlined above. Conversely, for the quantum case, we want
to show that if $\ultramodular$ is not ultramodular, then for
some quantum probability measure $\mu$, the composite
$\ultramodular\circ\mu$ might not be a QIVPM. Suppose there are three
particular numbers~$x_{0}$, $x_{1}$, and $x_{2}\in\left[0,1\right]$
such that $y=x_{0}+x_{1}+x_{2}\in\left[0,1\right]$, but they don't
satisfy Eq.~(\ref{eq:iota-Inclusion}). Consider the state:
\[
\rho=x_{0}\proj{0}+x_{1}\proj{1}+x_{2}\proj{2}+\left(1-y\right)\proj{3}\,.
\]
The induced map $\ultramodular\circ\muB_{\rho}$ constructed using the Born rule and
blurred by $\ultramodular$ fails to satisfy
Eq.~(\ref{eq:QuantumInterval-valuedProbability-Inclusion}) when
$P_{0}=\proj{0}+\proj{2}$ and $P_{1}=\proj{1}+\proj{2}$. In other
words, this induced map fails to be a QIVPM. For the classical case, if
$\ultramodular$ is not ultramodular, we also want to find a classical
probability measure $\mu:\eventsC\rightarrow\left[0,1\right]$ such
that $\ultramodular\circ\mu$ is not a classical IVPM. This can be
done by restricting our previous quantum probability measure
$\muB_{\rho}$ to the space of events  $\eventsC$  generated by the
mutually commuting projectors $\proj{0}$, $\proj{1}$, $\proj{2}$, and
$\proj{3}$.  The restricted function
$\mu=\muB_{\rho}|_{\eventsC}$ is then a classical probability
measure, and the induced map $\ultramodular\circ\mu$ fails to be
a classical IVPM for the same reason as in the quantum case.

In other words, essential properties of QIVPMs constructed using
interval maps can be gleaned from the properties of
ultramodular functions. The following is a most interesting property
in our setting:

\begin{thm}[Range of Ultramodular Functions]\label{thm:convex-uncountable}For
any ultramodular function~$\ultramodular:\left[0,1\right]\rightarrow\mathscr{I}$,
either $\mathscr{I}=\mathscr{I}_{0}$ as defined in Eq.~(\ref{eq:3-value-intervals}) or
$\mathscr{I}$ contains uncountably many intervals.\end{thm}

Since $\ultramodular$ maps to intervals, we can decompose it into
two functions: its left-end and right-end, where
 $\left[\ultramodularL\left(x\right),\ultramodularR\left(x\right)\right]=\ultramodular\left(x\right)$.
By Eq.~(\ref{eq:iota-Inclusion}), the left-end function $\ultramodularL:\left[0,1\right]\rightarrow\left[0,1\right]$
is Wright-convex~\cite{Wright1954,RobertsVarberg1973,PecaricTong1992},
i.e., 
\[
\ultramodularL\left(y\right)+\ultramodularL\left(x_{2}\right)\ge\ultramodularL\left(x_{0}+x_{2}\right)+\ultramodularL\left(x_{1}+x_{2}\right)
\]
for three numbers~$x_{0}$, $x_{1}$, and $x_{2}\in\left[0,1\right]$ with
$y=x_{0}+x_{1}+x_{2}\in\left[0,1\right]$. 
Together with the fact that $\ultramodularL$ maps to a bounded
interval $\left[0,1\right]$, the left-end function~$\ultramodularL$
must be continuous on the unit open interval
$\left(0,1\right)$~\cite{MarinacciMontrucchio2005}.  Therefore,
either $\ultramodular$ maps every number in $\left(0,1\right)$ to the
same interval, or the number of intervals to which $\ultramodular$ maps
must be uncountable.

To summarize, a conventional quantum probability measure has an
uncountable range $[0,1]$. A QIVPM constructed by blurring such a
conventional quantum probability measure must also have an uncountable
range of intervals. Of course, any particular QIVPM, or any particular
experiment, will use a fixed collection of intervals appropriate for
the resources and precision of the particular experiment. 

\section{Conclusion}
\label{sec:Conclusion}
  
Foundational concepts in quantum mechanics, such as the Kochen-Specker
and Gleason theorems, rely in subtle ways on the use of
unbounded resources. By assuming infinitely precise measurements,
these two insightful theorems form the foundations of two fundamental aspects of
quantum mechanics. On the one hand, the Kochen-Specker result reveals a
distinctive aspect of physical reality, the fact that it is
contextual, and, on the other hand, Gleason's theorem establishes a
relationship between quantum states and probabilities uniquely defined
by Born's rule.  Our goal in this paper has been to analyze the
physical consequences of a mathematical framework that allows for
finite precision measurements by introducing the concept of quantum
interval-valued probability. This framework incorporates uncertainty
in the measurement results by defining fuzzy probability measures, and
includes standard quantum measurement as the particular instance of
sharp, infinitely precise, intervals.  

In addition, we showed how these two theorems emerge as limiting cases
of this same framework, thus connecting two seemingly unrelated
aspects of quantum physics. We noted that {\it arbitrary\/} finite
precision measurement conditions can nullify the main tenets of both
theorems. However, by carefully specifying experimental uncertainties,
we were able to establish rigorous bounds on the validity of these two
theorems. Therefore, we have established a context in which infinite
precision quantum mechanical theories can be reconciled with finite
precision quantum mechanical measurements, and have provided a
possible resolution of the Meyer-Mermin debate on the impact of finite
precision on the Kochen-Specker
theorem~\cite{PhysRevLett.83.3751,Mermin1999}.

\section*{ACKNOWLEDGMENTS}

\begin{acknowledgments}
We gratefully acknowledge helpful comments from the anonymous reviewer that
greatly contributed to improving the final version of the paper.
\end{acknowledgments}

%

\end{document}